# 6. Malicious earworms and useful memes: How the far-right surfs TikTok's audio trends

Marloes Geboers and Marcus Bösch
Forthcoming in: Content Moderation Across Social Media Platforms, Ed. Richard Rogers

## Abstract

With its easy-to-use features of remix TikTok is the designated platform for meme-making and dissemination. Creative combinations of video, emoji, and filters allow for an endless stream of memes and trends animated by sound. From the get go, the platform focused its moderation on upholding (physical) safety, hence investing in the detection of harmful challenges. In response to the DSA, TikTok implemented opt-outs for personalized feeds and features allowing users to report illegal content. At the same time, the platform remains subject to scrutiny. Centering on the role of sound and its intersections with ambiguous memes, the presented research probed right-wing extremist formations relating to the 2024 German state elections. The analysis evidences how TikTok's sound infrastructure affords a sustained presence of xenophobic content that is often cloaked through vernacular modes of communication. These cloaking practices benefit from a sound infrastructure that affords the ongoing posting of user-generated sounds that instantly spread through the 'use-this-sound button'. Importantly, these sounds are often not *clearly* recognizable as 'networkers' of extremist content. Songs that do contain hateful lyrics are not eligible for personalized feeds, however, they remain online where they profit from intersecting with benign meme trends, rendering them visible in search results.

## Introduction

Toward the end of 2024, TikTok released updates to its community guidelines, accompanied by a brief acknowledgment of the difficulties that the short-form video format and the memetic logic built into the platform extend to moderation practices. The conjunction of users uploading their audios and the easy replication of sound afforded through the 'use this sound' button construct an environment where right-wing extremist actors easily surf on the vibes and rhythms of hateful audio memes. While sounds are deplatformed, they get ample time to prime or inspire others into re-uploading spinoffs of the same or an adjacent audio, rendering moderation especially cumbersome. As per TikTok's public announcement: *"Hate speech can be conveyed through any form of expression, including images, text, audio, cartoons, memes, objects, gestures, and symbols, some of which won't always be obvious."* The use of technology and human moderators to detect and remove such content is stipulated, mentioning for the first time how indirect attacks through jokes, memes, and audio trends will also be scrutinized by TikTok's moderation team. This includes any coded messages used by *"hate groups to communicate with each other without appearing hateful, such as code words, symbols, or audio trends."* This is quite an ambitious statement, as coded language and so-called algospeak are notoriously fluid, polysemous, and subject to change, and therefore hard to moderate (Steen et al., 2023). Our empirical analyses will make distinctions between overt hate speech across the several modalities of posts, and what we deem 'cloaked hate' to assess TikTok's moderation performance during the runup and aftermath of state elections held in Thüringen and Sachsen Anhalt on September 1, 2024. The conjunction of these elections and a fatal knife attack in Solingen on August 23 of that year spurred support for far-right extremist ideas embedded or otherwise linked to the political campaign of the Alternative für Deutschland (AfD) party. Especially in younger audiences, extremist ideas were shared and amplified by accounts on TikTok that participated in spreading far-right propagandist messages. In a report by the ISD (Institute for Strategic Dialogue, 2024) it was found that Neo-Nazis and white supremacists were sharing Hitler-related propaganda.

Failing transparency under the DSA

The presented research addresses how TikTok's sound architecture affords the proliferation and obfuscation of problematic content. Our research identifies loopholes that are exploited by malign actors, either surfing on benign audio trends or linking racist songs to popular meme templates. Moreover, we show how, even after time intervals of several weeks, problematic posts are only marginally moderated. Here, it is important to stress that we could merely determine if posts were no longer available, set to private view, or we could see that the account was no longer online. This does not necessarily imply platform moderation efforts, as users can take down posts and accounts as well. The platform only communicates that content or an account is no longer 'available. Under the DSA, TikTok publishes so-called Statements of Reasons, but in these vast spreadsheet documents, there is no identifier of the posts that were taken down by the platform. This effectively obstructs researchers from cross-checking whether unavailable posts were actually deplatformed.

During the COVID-19 pandemic and thereafter TikTok grew into a billion-user app with at the time of writing approximately 1.58 billion monthly active users worldwide. The multimodality of video posts affords easy disguising of messages as meaning can reside in a myriad of combinations of filters, emoji, hashtags, sounds, and gestures. This 'in-betweenness' makes far-right posts highly ambiguous, and therefore harder to read across vernacular cultures, after all, who actually knows that videos of motorbikes from the GDR-based brand Simson can signal a far-right audio meme that carries blatant racist messages in its lyrics? TikTok's nuanced nature significantly hampers moderation because: *"viral content that multiplies through memetic means with enough speed, anonymity, randomness, as well as some kind of logical order [..] can look organic or at least "semi-organic", whether or not it was originally a manipulated information or propaganda campaign"* (Galip, 2023, p. 101). Galip rightly points to this kind of 'covertness' as a phenomenon that benefits states and actors who hide behind a 'cloak' (Daniels, 2009) that gets drawn up as soon as users start mimicking a far-right TikTok meme. A targeted propagandist state-led injection of a racist message on this memetic platform, will, adopted by others, quickly become disconnected from its original instigator(s), allowing extremist messages to travel further through shapeshifting into a myriad of assembled posts, that repeat but also deviate in visual style or sound etc. On top of the affordances of multimodality, TikTok boasts a communicative infrastructure that entangles posts on various levels. Through encountering a 'regular' AfD (Alternative for Germany party) post displaying the speech of a politician calling for remigration (typically something a platform would leave untouched under the premise of free speech) a hashtag or a sound embedded in that 'benign post' links the artifact to wider, much more extremist, ephemera. Driven by the vibes of a sound, users can easily replicate the sound of someone else's post through the click of a button, and if problematic sounds are banned, there are thousands of potential users that can repost a sound under another name.

Rhizomatic soundscapes: TikTok's distributed audio infrastructure

While the app's short video format has permeated the social media landscape far beyond TikTok, the communicative architecture of said app is more or less 'unique'. The platform hosts a sound library with platform-listed sounds, however, the majority of users either use their own (pre)recorded sound or, even more ubiquitous: replicate someone else's sound. On the first occasion, users upload a video with a blend of audios that through this blend are 'unique'. These sounds contain voiceovers, music, speech, or sounds. In other cases the audio itself can be an exact copy of an existing song or sound, that is simply not taken from the library or not present as a listed sound. Both 'blended audio', as well as unlisted sounds are automatically registered as an 'Original Sound - [user name]' by the platform once a post is uploaded. If 'catchy enough', sounds take off through other users who encounter the post and use the 'use this sound' button. This effectively indexes their posts on a "sound page". When anyone clicks on the hyperlinked sound, they arrive at a collection of posts that use the same sound. A rather ubiquitous commenting practice is to ask creators the name of the song that was used. Such comments point to a desire to remix the sound with other audio such as problematic speech, birthing new soundscapes that are tethered to the original. In this way, sounds create spaces where actors find each other based on a shared sentiment (Papacharissi, 2015) and where they connect through e.g. participating via the 'use this sound' button or through remix, propelling fascist messages forward.

The networking affordances of sounds surpass hashtags in their intricate workings: sounds not merely gather posts that use the sound, there are also multiple versions of the same sound or song that are audibly similar, but that were uploaded by dispersed accounts and replicated by others. This creates distributed 'niches' or rhizomatic instances of what Geboers & Pilipets (2024) dub 'soundscapes': environments of creative assembly where various more or less homogeneous narratives can be hosted, 'held together' by the affective connotations that the sound in question evokes in its audiences. One can imagine how such distributed instances of (almost) the same sound severely hamper the platform's sound moderation. Once a sound comes to stand in for a hateful ideology, problematic accounts proliferating that ideology might be moderated, shadowbanned, or deplatformed, potentially also deplatforming any original sounds uploaded by those accounts, however, other instances of the same or a similar sound remain available. These practices include reusing sounds for coordinated campaigns, creating audio meme templates for rapid amplification and distribution, and deleting the original sounds to conceal the orchestrators' identities (Bösch & Divon, 2024).

The presented study takes as its focus the networked linkages between the AfD campaign and extremist rightwing niches on TikTok, 'held together' and networked through sounds that hijack popular songs or songs that are blatantly problematic in their contents but that latch onto meme templates to attain visibility. Attaching the political message of 'remigration' not merely to hashtags like #heimatliebeistkeinverbrechen (Love of home is not a crime) and #deutschejugendvoran (German youth forward), but also to sounds such as Gigi d'Agostino's *l'Amour Toujours* (the 'Gigi-song' hereafter), created sticky associations between upbeat club songs on the one hand, and the far-right political message of 'die Blauen' (blue is the color of the AfD with fans posting blue heart emojis) on the other hand. The online dynamic of co-opting the Gigi-song spilled over into a physical event where a group of young Germans on the terrace of Pony Club on the German island of Sylt, chanted the xenophobic slogan "Germany for the Germans, foreigners out" to the melody of the Gigi-song on Pentecost Sunday 2024. News media amplified the event, inadvertently recreating a new sound on the platform, to which, in turn, several TikTokers 'responded' by either replicating the sound through the 'use this sound feature' or by adding their own recorded file, establishing multiple soundscapes that all host the party crowd of Sylt. Alongside many malicious uses of the classic version of the

Gigi-song, remixes of Gigi d'Agostino–slowed, sped up, distorted, or otherwise tampered with–also connected those aiming to proliferate the AfD-led message of remigration. Alongside the Gigi-song, our study was able to detect linkages between far-right accounts and other trending sounds–such as the *'Kiss me'* sped-up version and *'Around the world'*–as well as sounds that harbor deeply problematic and violent content within their lyrics.

## TikTok & Content Moderation

From its launch in 2017, TikTok's community guidelines addressed hate speech and hateful behavior as prone to platform moderation. Despite this, the platform initially emphasized monitoring and deplatforming posts that are potentially risky for users' physical safety, with increasing attention to mental health throughout the years (Christin et al., 2024, OpenTerms Archive). What was deemed hate speech and hateful behavior was further specified after TikTok's surge in popularity in 2020, where the platform had to deal with emergent problematic phenomena exemplified by the rise in so-called QAnon accounts that propelled a popular conspiracy theory fuelled by pandemic-induced uncertainties. In October 2020, the platform announced a significant update to its policies on hate speech and misinformation, the company would harden its actions against QAnon initially merely banning specific hashtags as search terms. From then on, users sharing QAnon-related content on TikTok could expect their accounts to get deleted from the app. QAnon-associated search terms were redirected to a community guidelines warning. To give an example for current far-right spaces, the search term "Save Europe" and its associated hashtag, are also blocked. However, sounds and accounts boasting "Save Europe" in their titles or as account names were up and active during our research.

In March 2023, an update to the community guidelines was announced. Here, the emphasis was placed on guidelines for synthetic (AI-generated) media as well as on affirming the extant focus of TikTok when it comes to content moderation: safety from physical harm. From the get-go TikTok has placed the safety of minors at the forefront of their moderation strategies, extending warnings on posts that perform activities that potentially get someone hurt. Their content moderation rules dating from March 2023 clearly outline where the platform's priorities lie. Explicitly mentioned as main categories are youth safety and well-being, safety and civility, mental and behavioral health, sensitive and mature themes, integrity and authenticity, regulated goods and commercial activities, and privacy and security. Under 'safety and civility' hate speech is mentioned as a subcategory in the guidelines themselves, but there was, at the time of writing, no mention of hate and extremist ideologies in the widely distributed diagram itself.

While the initiative of warnings on posts displaying potentially dangerous activities is of course positive, the focus on physical safety comes with blind spots for other problematic forms of content. In the dataset used for the presented research, one post is particularly exemplary for the friction that arises when warnings are added to potentially dangerous activities whilst the post is surrounded by the extremist captions, sounds, and comments referencing clearly visible and common symbols of white supremacy. This post was networked through the hashtag #1161, which is a well-known numeric sign for fascists as the numbers translate to the letters AAFA (Anti-Anti-Fascist Action) and their position in the alphabet. Moreover, it uses an 'original sound' that serves as a networked space for an extremist community that meets through replicating these sounds. TikTok's focus on keeping people from imitating dangerous activities, such as smashing a bottle on the streets as displayed in the post itself, culminates in the bizarre situation where the posts will carry a warning relating to the smashing of a bottle while leaving other highly problematic content elements in that post and in its comments unaddressed.

Throughout its years of existence, the most prominent development in relation to community guidelines addressing hate speech and hateful behavior was the further specification of so-called categories of hate. These would include claims of racial supremacy, misogyny, anti-LGBTQ+, and antisemitism. Islamophobia was added later to the list (September 2024). While TikTok outlines how the app uses a myriad of automated technologies–ranging from text-based NLP approaches to computer vision algorithms–to scrutinize posts on various modalities (images, account names, sound, descriptions, gestures, symbols, etc), earlier research found that TikTok moderates in generally light and highly inconsistent ways (Zeng & Kaye, 2021). There is a growing awareness that its multimodality urges researchers to include audiovisual layers of 'user-led obfuscation' (Bösch & Divon, 2024). Turning trending songs into problematic content or associating trends with hateful conduct is a strategy of 'creatively' obfuscating worrisome materials. For example, early into the Russian invasion of Ukraine, the song *Good Evening Ukraine* was turned into *Good Morning Russia*, surfing on the popularity of the original (Pilipets et al., 2025), copying the tune and simply replacing the refrain. While, after some months, the Russian version and its soundscape turned out to be unavailable, there were still a handful of other soundscapes carrying the same sound. This problem directly relates to the inherent affordance of the 'original sound' that through the 'use this sound button' enables fast replication of audios that were initially uploaded by problematic actors. While the initial 'soundscape' instigated by a problematic actor might get banned, many others heard the sound before the account was deleted. These 'sound-followers' will then re-create their own original sound, that sounds exactly the same.

*Engagement over well-being*
It is clear that audio presents a fundamentally different set of challenges for moderation than text-based communication (Brisset, ADL). While the platform's official communication outlines efforts and investments undertaken to enhance better (sound) moderation technologies, and while recently there has been an acknowledgment of the presence of indirect or 'less obvious' hate as subsumed in memes and audio trends, these developments merely point to growing awareness, and not so much to actual enforcement of the app's own community guidelines and its moderation framework. Echoing Christin et al. (2024) and their so-called 'internal fractures of social media companies', the pattern seen in moderation of far-right accounts is signaling how the winning logic for platform companies is not that of user well-being, but that of engagement. Researching a variety of platforms, including TikTok, Christin et al. (2024) outlined some vignettes based on leaked information, that evidence the competing logics and how engagement overrules moderation. One of these vignettes was the retraction of an algorithm developed by Facebook employees in late 2020: When [they] realized that their users viewed many of the most viral posts on the platform as "bad for the world", a machine learning classifier was developed to downrank such posts, only to have the effort shelved by executives because it reduced engagement metrics. (Roose et al., 2020). Christin et al.'s hypothesis that the logic of engagement over well-being would also hold true for TikTok was affirmed in a 2024 case where the ISD reported 50 accounts for far-right hate speech to test how TikTok would moderate them, while the platform eventually removed violative content and channels, in many cases it does so only after accounts have had the opportunity to accumulate significant viewership. For example, the 23 banned accounts from the sample dataset managed to accrue at least 2 million views across their content prior to getting banned."

*How niche soundscapes sustain hate*
The presented analysis addresses the role of problematic 'original sounds' in the prolonged existence and proliferation of far-right white supremacist and xenophobic content on TikTok. We charted how far-right actors co-opt trend visibilities (and meme templates) on the platform and assessed the presence of problematic content in For You Feeds across three countries. Through moderation (or disappearance) trace analysis of videos and accounts that went offline after intervals of some months, we laid bare how so-called 'niche soundscapes of hate' are either very lightly moderated and/or disappear from the platform due to user-led decisions of deletion. To conclude, we will connect findings to a continuation of the platform's emphasis on preventing physical harm, a reliance on crowdsourced (user-led) flagging practices that only emerge when posts achieve a significant reach, and a demotion of niche hateful content, that, while posts are ineligible for the feed, still prove able to maintain and connect communities that can converge over extremist ideas. When smartly attached to current trends on the platform, such niche audio memes can appear in people's search results when looking for trending sounds. To summarize, our research investigates the following questions: 'How is TikTok's sound architecture affording the proliferation and obfuscation of problematic content?' And to what extent are problematic posts moderated or visible?

## Methods

TikTok is built around easy imitation where users respond to each other through replicating and remixing sounds, hashtags, emoji, filters, and so on (Zulli & Zulli, 2022). Our methods are attuned to charting the tactical practices of malign actors that latch onto benign (audio) trends on TikTok through imitation, in order to amplify as well as prolong the 'lives' of far-right messages. The first part of our empirical assessment of the presence of extremist content on TikTok centered on the user experience of their personalized feed. While the search feature is increasingly used–AI Forensics (2023) found that 67% of TikTok users in Germany regularly use the search function–it is not possible to determine to what extent extremist posts are recommended to users on their For You Feeds. By merely assessing browser-based search results, we maintain a blind spot for the ways in which people encounter right-wing extremist content in their feeds. Therefore, we conducted a persona-based analysis (Bounegru & Weltevrede, 2022) where we assess how a right-wing person in Germany might be confronted with borderline or downright fascist content on their feeds, and how the presence of borderline or problematic content in their recommendations compares to rightwing users in the Netherlands and in the UK. As the authors were located in both Germany and the Netherlands we merely had to set up clean accounts using a developer browser that deletes cookies and traces whenever a browser session is closed. For the UK feed we used VPN software. Personas were 'pre-trained' to signal rightwing interest, through following corresponding political party accounts and conservative news outlets.

*Hijacking trends and cloaking hateful lyrics*
After establishing the presence of extremist content in feeds of rightwing users across three countries, our second step is to assess the longevity of the 'lives' of extremist audio memes: are they moderated, or at the very least: do these audio-driven memes go offline after some time has passed? Using the mobile app search feature we mirrored a user actively searching for events such as the Solingen knife attack of August 2024 (Light et al, 2018). Two prominent hashtags present in this 'walkthrough observation' were #deutschejugendvoran and #heimatliebeistkeinverbrechen. These hashtags were used as search queries to collect posts and metadata using the DMI-tool 'Zeeschuimer' (Peeters,

2021). This tool outputs data on the authors of posts, follower counts, sounds used, timestamps, hashtags, overlaid texts, filters, and a range of engagement metrics such as likes and plays.

Through a systematic assessment of the metadata of posts using one of the two problematic hashtags, we laid bare patterns derived from so-called 'sticker texts': the overlaid texts on top of videos. Recurring texts point to the presence of meme templates or trends. One of such trends found in our hashtag space was *'Music taste is important, imagine you do not know what comes after the intro'*. This trend hinged on users playing the intro of a song and then letting others comment on the song's name and sometimes (the rest of) the lyrics. Outside of our dataset with nationalist hashtags, this 'trend space' harbors songs that are completely benign, but our hashtagged posts made use of the trend by playing the intro of a song we will here dub *Türke*. The lyrics are (partly) cited in the comments section. When we searched with a clean account for '*Musikgeschmack ist wichtig*', the first malign use appeared as the 9th search result. The lyrics are derogatory toward people of Turkish descent, it humiliates people and calls out to kill them. The Music taste-trend lends itself perfectly for what we call a 'cloaking while amplifying' strategy. Playing merely the intro, without lyrics, works as a dog whistle for those in the know. A detection algorithm will not find the derogatory lyrics as they are not in the post and oftentimes also not (fully) present in the comments.

A second song–or audio meme–found in the hashtag-based dataset was *Zecken* (translates as ticks). This sound distinguishes itself from the *Türke* song, as it puts its racist lyrics on full display, where immigrants are compared to ticks and are punched to death. TikTok states that: *"Dehumanizing someone on the basis of their protected attributes by saying or implying they are physically, mentally, or morally inferior, or calling them degrading terms, such as saying they are criminals or animals, or comparing them to inanimate objects"* is against guidelines. The song hinges on a peculiar remix between a folk song named *Kreuzberger Nächte sind Lang* birthed in the 1970s, and a 1997 song by the Zillertaler Türkenjäger that blended the Kreuzberger refrain with racist lyrics where the verses outline how *'Zecken'* (leftists and punks are all ticks in vernacular language) and 'Kanaken' (a German ethnic slur for people with roots from Southeast Europe, Middle East, and Northern Africa) are encountered on the streets of Kreuzberg (Berlin), and are then violently 'ended'. When searching for this song with a clean account in August of 2024, the first two posts contain the toxic version with racist lyrics. Interesting is how an AfD fan account hosting 111.000 followers posted a video of a 1970s performance of the innocent song on September 19, 2023, close to the start of the AfD campaign. It might have primed extremist TikTokers to use the vile remix of 1997.

Following the approach of 'moderation trace analysis' (De Keulenaar et al., 2023) we probed the online status of the post links in early October 2024 (two months after the initial collection of data), and again in early December to assess what posts 'disappeared from public online view'. As mentioned, due to a lack of transparency, we are, unfortunately, not able to cross-check whether offline posts were taken down by their creators or by the platform. We could distinguish between videos that over time turned to 'unavailable', or were 'set to private', and we could, through the post URL of unavailable videos, detect accounts that were deleted. At least in the case of the deleted accounts we can assume that many of these point to instances of deplatforming rather than users voluntarily deleting their 'social capital'.

*Disgust and contestation: network analysis of sounds and accounts*
While largely seeking to recruit the like-minded, memetic engagement on TikTok does not exclude contestation (Geboers & Pilipets, 2024). When sounds catch on and users respond with novel adaptations: *"positive, negative, and ambivalent affect blend into each other"* (Paasonen, 2019, p. 52), constructing an environment of oscillating affective charges. It is this oscillation between excitement and disgust that keeps engagement with political messages 'on the move'. This movement is a prerequisite for staying relevant in the ephemeral spaces of the platform, as well as for the consolidation of political extremist messages. This led us to perform a network analysis that charts the relationships between accounts and sounds within a space of fascist and anti-fascist contestation. This network was based on data collected through a hashtag-based query using #1161, a known numeric symbol for fascists. This hashtag is often countered by posts using #161 which stands for anti-fascist movements. We color-coded accounts for the presence of political extremist content, which materialize on the level of the hashtags (#1161 or #88 (for "Heil Hitler") and so on, using the ADL Hate Database) but which can also sit in the performance of gestures, the depiction of symbols such as the 'Black Sun' or Sonnenrad (sun wheel) which is communicated through the spiderweb emoji and by depicting and referencing so-called 'Talahons', adherents of an urban subculture of males, typically but not necessarily of Middle Eastern origin, characterized, among other things, by a passion for German hip hop and wearing counterfeit designer labels. This caricature has been charged with ever more negative connotations, in part through the viral AI-generated song *Verknallt in ein Talahon* (Crush on a Talahon), which details how a white young woman falls in love with a criminal foreigner.

Sounds were color-coded for hosting extremist content on the level of their audios. The size of the triangle nodes (sounds) was based on the number of posts that used that sound, providing an idea of the popularity of that particular sound. The size of the circle nodes (accounts) was based on the number of followers of an account. We coded for the presence of fascist content based on the account's posts present in our dataset as well as on their larger profiles. Account names were replaced by numbers in the network.

In a last step, we selected one of the problematic sounds found in the #1161 dataset, we engaged in querying the sound's name (*Anotha Europe*) in the mobile app's sound search module. The accounts found in this search either proved to be authors of sounds that resemble closely *Anotha Europe*, or, even more interestingly, use remixes, that blend the sound with various versions of benign hit songs. The output of this analysis is a linear dendrogram (Figure 6.3) depicting prominent accounts and popular sounds they 'authored' (uploaded to the platform).

## Findings

To get a sense of the visibilities of far-right content in the personalized For You Feed (FYF), we trained rightwing personas (newly set up accounts) and used VPNs to emulate the experience of the users across three countries when scrolling through their feeds. We mapped the presence of right-wing extremist posts (red in Figure 6.1) as well as borderline posts (orange in Figure 6.1) in the feeds of these personas across countries. We can see how the German feed shows slightly more problematic and borderline posts on TikTok as compared to the feeds of right-wing users located in the UK and the Netherlands. To what extent this discrepancy is related to the campaign-related activities of the German AfD party, is hard to establish, but there is no doubt about the effectiveness of its sound-based propaganda tactics. The party and their fan army in a participatory propaganda approach play the algorithm to ensure-far reaching pro-AfD content (Bösch, 2023) "flooding TikTok" ([Breschendorf, 2024](#)) with songs and sounds.

Interestingly, TikTok has a separate section where it outlines a number of content characteristics that will turn posts 'not eligible for the FYF'. TikTok's community guidelines state that: *"Content may be ineligible for the FYF when it indirectly demeans protected groups.* [..] Protected groups means individuals or communities that share protected attributes such as race and gender. These attributes include immigration status and national origin.

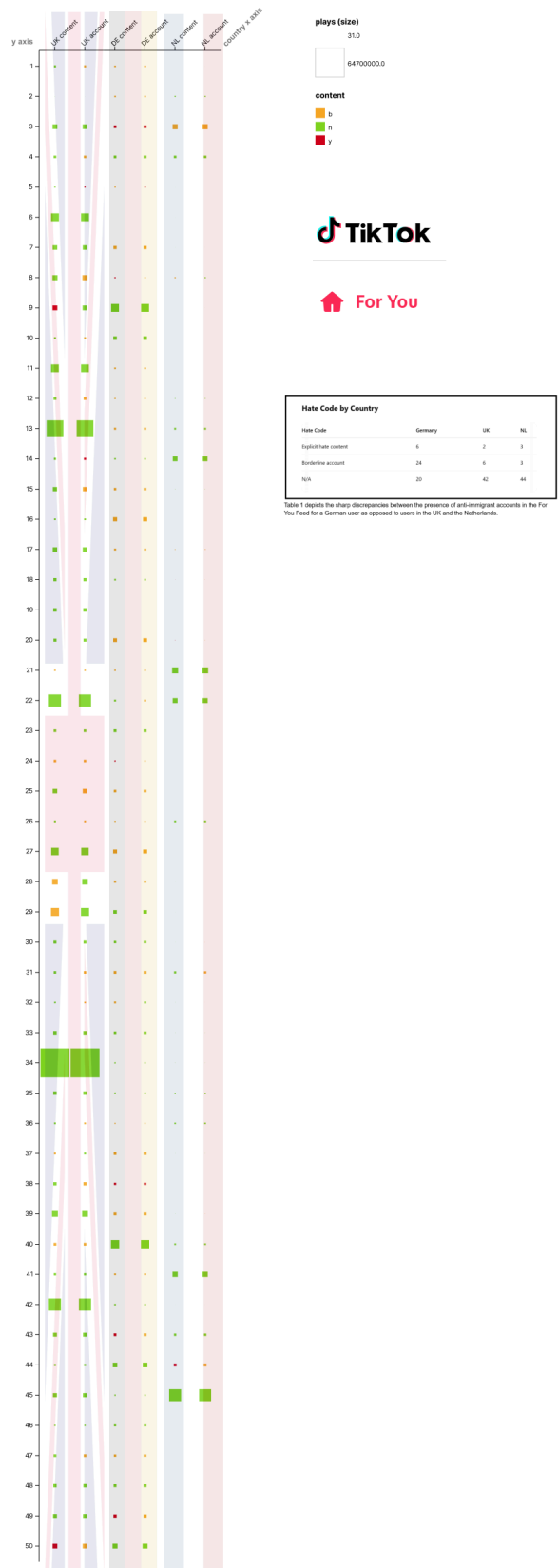

| Hate Code by Country | | | |
|---|---|---|---|
| Hate Code | Germany | UK | NL |
| Explicit hate content | 6 | 2 | 3 |
| Borderline account | 24 | 6 | 3 |
| N/A | 20 | 42 | 44 |

Table 1 depicts the sharp discrepancies between the presence of anti-immigrant accounts in the For You Feed for a German user as opposed to users in the UK and the Netherlands.

**Figure 6.1** displays the presence of extremist (red) and borderline (orange) posts in the feeds of trained accounts across the UK, Germany, and the Netherlands. Borderline posts are posts not directly carrying explicitly hateful content, but that are connected to hateful anti-immigration accounts. The numbers in the inserted table depict the sharp discrepancies between the presence of anti-immigrant accounts in the For You Feed for a German user as opposed to users located in the UK and the Netherlands. Source: authors.

*The role of fascist and anti-fascist contestation*
To address the question of the role of TikTok's networked sound infrastructure within spaces of political contestation, we gathered data collected through the hashtags #1161 (anti-anti-fascists) and #161 (anti-fascists). The network shows how accounts (kept anonymous through representing them as numbers) connect to various more or less popular and more or less frequently replicated sounds. Green accounts are accounts countering the #1161 message, red nodes represent fascist accounts. Account nodes (circles) are sized by their amount of followers. Sounds are represented as triangles and colored purple for mainstream audio trends on TikTok and brown for niche fascist songs. Their sizes correspond to the number of posts that copied the sound. While there are brown songs, the majority of fascist accounts assemble around mainstream, trending songs such as a sped-up version of *Kiss me,* the already mentioned Gigi-song (*l'Amour Toujours*), and *Around the World (lalala).* Interestingly, an anti-fascist account with 21.900 followers, uses the much more niche fascist sound of *Scheiss Egal* to purposely target fascists.

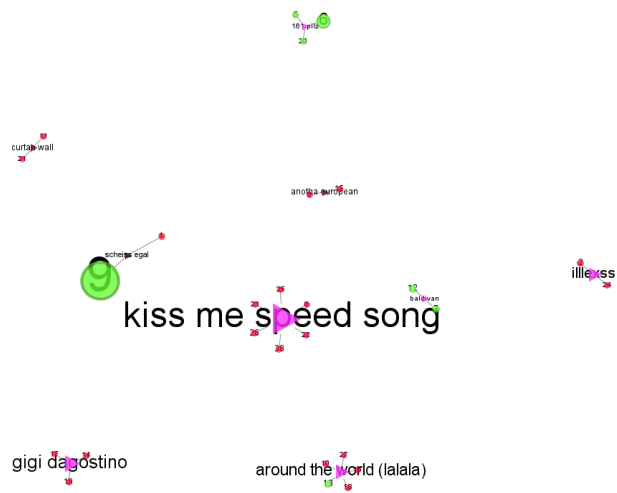

**Figure 6.2** displays anonymized accounts (numbered), their fascist or anti-fascist stances in color (green for anti-fascists and red for fascists), and their connections to sounds that were either benign in audio content but hijacked by extremists on the platform (purple triangles), or that were fascist and racist in their audio contents (brown triangles). Source: authors.

Note that in the network the sound nodes represent multiple similar 'original sounds' that were all uploaded to the platform separately. Some of the rightwing *Scheiss Egal* sound versions were at one point no longer available which points to their creators (initial original sound uploaders) being deplatformed, or they have purposefully deleted their own accounts. Other versions remained online, pointing out the complexity of effective moderation in a landscape of replication.

*Soundscapes as dispersed problematic niches*
To further investigate how versions of similar sounds perpetuate and prolong the proliferation of fascist content, we turned to the mobile app sound search, where we queried one of the brown-colored 'niche' songs from the network in Figure 6.2. Querying '*Anotha European*' in the mobile app lets us manually assemble sounds and accounts that first uploaded these sounds. The app also displays the number of posts that replicated that particular sound. A set of eight accounts connected to songs that were all also present in our #1161 dataset. One account (Figure 6.3) was particularly interesting as its account name consists of a banned search query (Save Europe) and it ties into sounds that host extremist content (the Gigi-song, *Kiss me Speed, Around the World (lalala)* and a variety of Russian-language songs such as *Australia* (by the band called X) not depicted in the dendrogram. This highly active account hosts 19,000 followers despite its user name representing a banned search query. The sounds also show how they–modulated by memetic replication–morph into remixes that speed up, slow, or otherwise distort the original, perhaps being part of a user imaginary in which reworked sounds work to evade moderation.

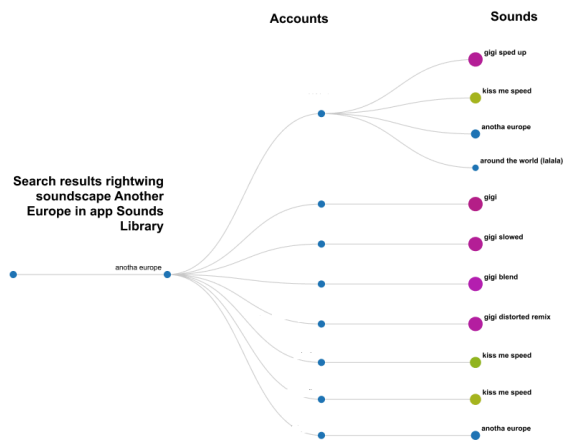

**Figure 6.3** displays mobile app 'sound search' results for the query 'Another Europe', which was based on a problematic song title found in the sounds-accounts analysis (Figure 6.2). The search results contain accounts that uploaded (versions of) both 'Anotha Europe' as well as the more pop culture sounds of 'Kiss me again' and l'Amour Toujours by Gigi d'Agostino. The colors point out audio similarities between the various remixes. The first account (on top of the middle row of nodes) engaged in multiple 'plugs' of sounds, all of them very familiar to rightwing vernacular tropes on TikTok. Source: authors.

The dendrogram of in-app search results (Figure 6.3) displays how the Gigi-song reverberates through spinoffs that speed up, slow down, distort, or blend in with other sounds and speeches. This inspired us to 'reverse our search' by querying the Gigi song *l'Amour Toujours.* This, disturbingly enough, showcased within the first 50 results, a sound that not merely subsumes the Gigi tune, it also blended it with a speech stating how 'Hitler, war der Mann ihres Lebens' (Hitler was the man of your life). This sound was replicated through the 'use this sound' button 1036 times. The post using this sound was, at the time of writing, banned or taken offline by the account owner. But the first post of the (still available) account was online and asked people if they believed that it 'really was 6 million', downplaying the death toll of the Holocaust.

*Moderation traces: minimal 'disappearances'*
To address the question of the extent of moderation taking place, we conducted a trace analysis. As mentioned, due to a lack of transparency on the side of the platform, as researchers we can merely determine whether videos or accounts are no longer available with no way of knowing whether content was taken offline by the TikTok user or by the platform. We nonetheless engaged in an assessment of the status of post URLs after two designated intervals after collecting posts toward the end of August 2024: early October and early December of the same year. We distinguish between videos that are offline, accounts that no longer exist, and accounts that are set to 'private'.

We selected two problematic songs that were present in the hashtag-based datasets for #heimatliebeistkeinverbrechen and #deutschejugendvoran. We used the indexed sound pages for these songs and ran the browser-based scraper tool Zeeschuimer to collect posts and metadata. We revisited post URLs twice, each with a two-month interval. The first song on TikTok 'merely' conveys the intro, no lyrics are heard, just the techno beat of *Türke*. The text taps into a meme trend that asks whether people recognize the song in a ludic way: *'Imagine you do not know what comes after the intro'*.

For assessing moderation dynamics in this soundscape, we engaged in analytically distinguishing between 1) posts carrying hate speech on the textual post dimensions of captions and overlaid video texts (so-called 'Stickers'), and 2) posts containing coded or ambiguous visual symbols signifying or connecting to extremist rightwing tendencies (see Figure 6.4). The left diagram in Figure 6.4 represents 65 Türke-posts that could be coded for four common and highly ambiguous (hard to read) memes present in the hashtag-based dataset. These memes showcase motorbikes of the East German brand Simson, tractors, working shoes as stand-ins for 'real workers' as opposed to the sneakers of immigrants, and lastly the 'hampelmänner' (puppet on a string) meme that symbolizes how you can either be led by others or do your own research on immigrants. Note that the puppet on a string memes seem 'better moderated', although this could also be a misleading image: this meme could have assembled accounts that were also engaging in other more explicit hate speech posts, hence their account deletions. We assessed URL status over time showing how 55 (October) and 49 (December) posts were still online which amounts to 85% and 75% of the total of 65 posts with hate cloaked as an ambiguous meme. There seems to be a slow and quite minimal decline of available posts in this ambiguous space which is to be expected when content is 'cloaked' to this extent. When compared to posts carrying explicit hate speech (right diagram in Figure 6.4, representing 38 posts) in textual layers of *Türke* posts, we see that 66% (October) and 53% (December) of these posts were still online. Figure 6.4 allows us to see how memes without textual references to hate are indeed much less moderated than posts that have some form of hate speech or hateful symbols and references to ideologies (gestures, numerical signs) subsumed in their textual layers.

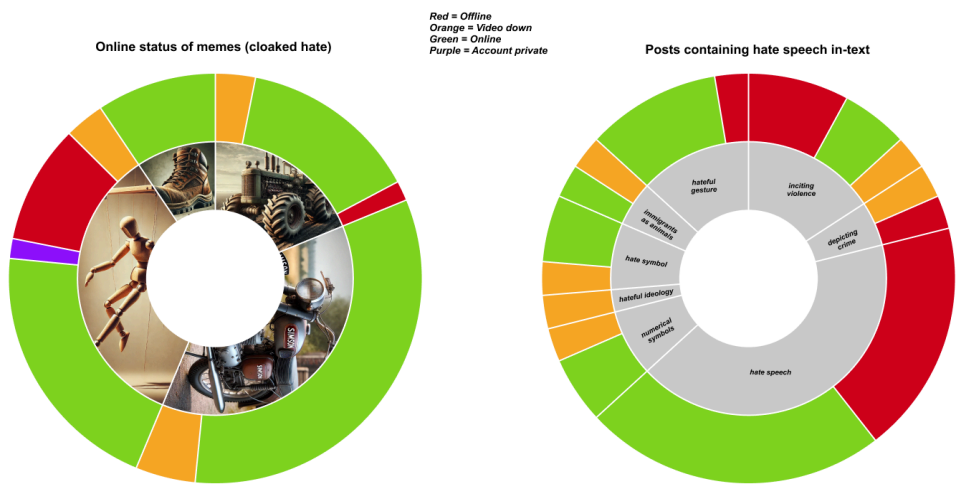

**Figure 6.4** displays the status transformation of post URLs between the time of data collection (August 2024) and revisiting these URLs early December 2024. Both diagrams depict posts using the intro tune of the xenophobic Türke song in combination with the 'Musik geschmack ist wichtig' trend that materializes in the overlaid text on top of the videos (sticker texts). To the left we see the status of posts that were 'double-cloaked' using the intro tune of the hateful song and a particular visual template (recurring memes) that works as 'coded visual language for rightwing extremism). To the right we see posts using the same tune and sticker text, but these posts also carry explicit hate and violence in textual post layers, making them more recognizable as hate speech. The categories of textual hate are derived from TikTok's community guidelines. Source: authors.

The second song (Zecken, or Ticks) boasts explicit lyrics about killing immigrants on the streets of Kreuzberg, Berlin. Here, all posts on the sound page (or in the soundscape) carry explicit hate as the lyrics are in large part cited in the videos. To further assess how moderation takes place across modalities of audio and text, we decided to distinguish between posts carrying textual hate (as added to the explicit audio) and those that do not (figure 6.5) to see whether this influenced the online presence of such posts over time.

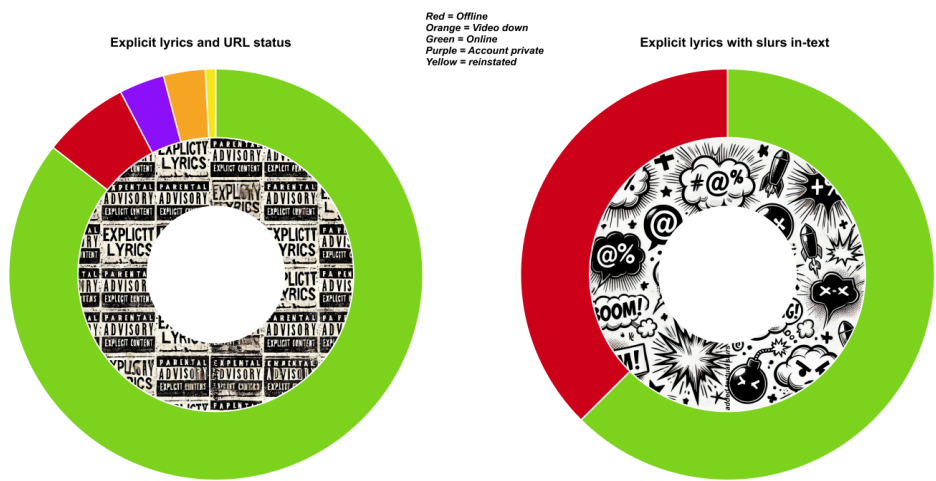

**Figure 6.5**
displays the status transformation of post URLs between the time of data collection (August 2024) and revisiting these URLs early December 2024. Both diagrams depict posts using the explicitly hateful song 'Zecken' (ticks as referring to immigrants). To the left, we see the entire dataset and its status transformation. To the right, we see the posts that express hate and violence in the textual layers. The categories of textual hate are derived from TikTok's community guidelines. Source: authors.

Despite the explicit presence of speech violating the TikTok community guidelines of hateful behavior, the level of moderation is comparable to the more covert dog whistle sound of *Türke*. In the overall dataset, 89% of URLs in the first and 86% in the second interval remained online, in the posts containing textual hate speech (left diagram in figure 6.5) these percentages dropped to 65% and 59% respectively. This points to textual hate as a greater predictor for the moderation (or at least 'disappearance') of posts than the presence of hate speech on the level of audio.

### Undercurrents of extremism: how hateful content circulates under the radar

TikTok affords easy imitation of audio content through the 'use this sound' button, allowing audio memes to go viral (Abidin, 2021). As such, a sound is made searchable and can harbor a wide array of topics. Very often though, posts networked by sounds, skew toward particular topics, including those deemed problematic. While TikTok's strategies for moderation is said to include audio (Hee et al.,

2024; Steen et al, 2023; Medina Serrano, 2021) the polysemous nature of sound, like images, hampers effective moderation. The multimodality of TikTok further complicates this matter, as a 'benign sound' can shift meaning quickly when used in tandem with particular textual or visual images. Moreover, as we laid bare in the outlined study, the infrastructure of original sounds that are easily replicated, creates connections between highly visible audio trends and malign actors that are close to impossible to moderate, as the sounds of banned accounts can continuously re-emerge under new versions of sounds.

We focused our analysis on sounds and so-called soundscapes as spaces where communities affectively convene around practices of sound-replication. To further exemplify how TikTok is the ideal platform for extending and prolonging extremist 'vibes', also on the level of (ambiguous) visual modalities, we want to outline one of the memes present in the Türke sound, that is the recurring motorbikes of a particular brand. The love for Simson motorbikes was most probably inspired by an AfD promotion poster showcasing politician Björn Höcke riding the signature East German Simson motorbike. This inspired many TikTokers to post their own motorbikes on the platform. From a study of similar motorbike memes connected to another far-right sound–'*Anotha Europe*'–Geboers et al. (2025) found that such motorbike memes pull in both far-right accounts as well as (many more) benign motorists who very probably do not have a clue that they are connecting their motorbike to a racist tune. From a sample of this study containing 36 accounts, the researchers found only three clearly fascist accounts, the other 33 accounts were not clearly rightwing. This unconscious connection between a hobby (motorbikes) and a political ideology subsumed in an ambiguous song intro, expands the affective reach of far-right sentiment beyond explicitly political spaces. When a 13-year-old with a cool bike picture slideshow sees other bike posts and decides to use the same sound, this person's post gets networked, through the sound linkage, to far-right users and their extremist content.

*MLLM-detection: not a fix for all*
With the development of MLLMs the detection of hate speech with more context sensitivity is around the corner (Hee et al, 2024). However, these advancements will push users to engage in more latently present hate speech. While algospeak is associated with misspellings, emoji, or interpunction tactics (Steen et al., 2023), sound introduces more 'subtle' tactics of circumvention. This project responded to a scholarly and institutional (corporate and political) need to recognize that algospeak is "more than the simple replacement of words [..] it needs to be understood as code words and linguistic variations, visual and multimodal communication, and audiovisual coherences [..]" (Steen et al., 2023, p. 12). Our study honed in on the audiovisual character of the platform and its infrastructure of networked sounds that create loopholes for politically extreme actors to stay active and garner levels of engagement that when we account for the accumulated engagement that dispersed original sounds assemble, are certainly not negligible.

From our trace analysis of moderation (or tracing disappearance) of videos and accounts in problematic soundscapes we found minimal 'deplatforming' or disappearance of hateful content, even when hate is explicitly present in either audio or text. In order to understand these minimal levels of moderation, we need to take into account TikTok's reliance on flagging practices. It is no coincidence that directly underneath the list of direct and indirect hate speech in the platform community guidelines, there is a section titled 'What do I do if I or someone I know experiences hate on Tiktok?' The app relies heavily on the practice of flagging, allowing users to report content that violates community guidelines (Crawford & Gillespie, 2016). Are describes how flagging facilitates liability evasion for platforms and a delegation of labor, but "its influence on moderation remains opaque"

(2024, p. 4). The rather minimal disappearances of problematic videos and accounts in our analyses might very well be related to the platform's reliance on users flagging content to platform moderators. Taking into account how sounds such as *Türke* and *Zecken* are circulating within niche and distributed soundscapes, most often not attaining high engagement metrics, we might derive that these posts are ineligible for the FYF. This also means that the chances they get flagged are close to zero. This would explain how such posts are largely neglected by moderators.

*Strategies of cloaking while amplifying*
Based on the empirical analyses, we were able to identify three strategies that users adopt to avoid moderation. We regard these strategies to follow the logic of 'cloaking while amplifying' extremist messages. These tactics are respectively: 1) creating sticky associations between far-right extremist and racist ideas under the 'cloak' of trending songs (such as putting to work a highly popular song of Gigi d'Agostino, 2) proliferating extremist audio (mainly songs, at times blended with problematic speech) by tapping into benign trending memes that play fragments of songs and intros (such as *Musik geschmack ist wichtig/Music taste is important*), and 3) remixing an innocent tune to map explicit lyrics onto these tunes (*Zecken* as remixed with the folk song *Kreuzberger Nächte sind Lang*). As said, the larger part of these posts do not boast significantly high engagement metrics, with a mean of 1821 plays for *Türke* and 1457 plays for *Zecken* posts. For comparison, the posts collected with #1161, which holds various racist but also popular mainstream songs, has a mean of 73,352 plays. Nonetheless, niche soundscapes do appear in search results when users query for a trend (*Musik geschmack*), or when users search for benign sounds that get remixed in with vile versions of the sound (*Zecken*). Thus, while possible demotion efforts might underlie the overall low engagement of these posts, the fact that these niche soundscapes are scarcely moderated means that they continue to sustain an extremist 'vibe'. They serve as spaces of connection that keep extremist communities alive in the undercurrents of the platform. Occasionally, their posts, when linked to popular visual memes (motorbikes) and memetic templates such as *'Musik geschmack ist wichtig'* do confront users with extremist beliefs.